\documentstyle[12pt,amssymb]{article}

\def\be{\begin{equation}}
\def\ee{\end{equation}}

\def\b{\hfill\break}

\title{
Motion of four-dimensional rigid body around a fixed point: an
elementary approach. I. }

\author{A.M. Perelomov
\footnote{On leave of absence from Institute for Theoretical and Experimental
Physics, 117259 Moscow, Russia. Current e-mail address:
perelomo@dftuz.unizar.es}\\
{\small\em
Departamento de F\'{\i}sica Te\'orica, Universidad de Zaragoza,}\\
{\small\em E-50009 Zaragoza, Spain }}

\date{}
\begin{document}
\maketitle{}

\begin{abstract}\noindent
The goal of this note is to give the explicit solution of
Euler-Frahm equations for the Manakov four-dimensional case by
elementary means. For this, we use some results from the original
papers by Schottky [Sch 1891], K\"otter [Koe 1892], Weber [We
1878], and Caspary [Ca 1893]. We hope that such approach will be
useful for the solution of the problem of $n$-dimensional top.
\end{abstract}

\setcounter{equation}{0}

\noindent {\bf 1.} The equations of motion for a rigid body in a
four-dimensional Euclidean space with a fixed point coinciding
with the center of mass (and also for the $n$-dimensional case)
are the generalization of famous Euler's equations. They were
found first by Frahm [Fr 1874]\footnote{The problem of
generalization of Euler's equations was posed by Cayley [Ca
1846].} and they have the form
 \be \dot l_{ij}=\sum _{k=1}^4(l_{ik}\,\omega _{kj} -\omega_{ik}\,l_{kj}),\quad
 \omega_{ij}=c_{ij}\,l_{ij},\quad l_{ij}=-l_{ji},\quad i,j=1,\ldots ,4.\ee
Here $c_{ij}=I_{ij}^{-1}$, the dot denotes the derivative with
respect to time $t$, and $l_{ik}$, $\omega _{jk}$, and $I_{ik}$
are components of angular momentum, angular velocity and principal
momenta of inertia tensors, respectively.

 In this paper we consider completely integrable
 Manakov's case [Ma 1977], when quantities $c_{ij}$ have the
form\footnote{\,Note that for the "physical" rigid body
$c_{ij}=I_{ij}^{-1}$, $I_{ij}=I_i+I_j$. In this paper we consider
a general integrable case when quantities $c_{ij}$ and $I_{ij}$
are arbitrary.}
 \be c_{ij}=\frac{b_{i}-b_{j}}{a_{i}-a_{j}}.\ee
 In a number of papers (see [AM 1982], [Ha 1983], [AM 1988],
 and references therein) so called method of linearization on
 the Jacobian of a spectral
curve defined by the characteristic polynomial of one of the
matrix in the Lax pair was used. However, as it was mention in [AM
1988], "this approach has remained unsatisfactory; indeed (i)
finding such families of Lax pairs often requires just as much
ingenuity and luck as to actually solve the problem; (ii) it often
conceals the actual geometry of the problem".

So, in the present note we return to the original
Schottky--K\"otter approach [Sch 1891], [Koe 1892]. In our
opinion, this elementary and natural approach is more adequate for
the problem under consideration. We hope that it will be useful
also for the more complicated problem of $n$-dimensional top at
$n>4$.

Let us remind that in the  paper [Sch 1891] the problem under
consideration was reduced to the Clebsch problem [Cl 1871] of the
motion of a rigid body in an ideal fluid\footnote{\,\,This result
was rediscovered one century later in the paper [Bo 1986].}. For
the special cases, the last problem was integrated explicitly by
Weber [We 1878] and by K\"otter [Koe 1892].

However, the Clebsch problem is related not to $so(4)$ Lie algebra
but to the $e(3)$ Lie algebra -- the Lie algebra of motion of the
three-dimensional Euclidean space. Hence, it is important to
extend the Schottky--K\"otter approach to give the solution in
$so(4)$ covariant form. Here we give such a solution using the
elementary means\,\footnote{\,A special $so(4)$ case with tensor
$l_{jk}$ of rank 2 was integrated explicitly by Moser [Mo 1980].}.

\medskip\noindent
{\bf 2.}  Note first at all that equations (1) are Hamiltonian
with respect to the Poisson structure for the $so(4)$ Lie algebra
-- the Lie algebra of rotations of the four-dimensional Euclidean
space,
 \be \{l_{ij},l_{km}\} =l_{im}\,\delta _{jk}-l_{ik}\,\delta _{jm}+l_{jk}\,
 \delta _{im}-l_{jm}\,\delta _{ik}. \ee
  The Hamiltonian is given by the formula
  \be H=\frac12\,\sum  _{j<k}^4 c_{jk}\,{l_{jk}^2}\,, \ee
 where quantities $c_{ij}$ are given by formula (2), and equations (1)
 may be written in the form
  \be \dot l_{jk}= \{H,l_{jk}\}. \ee
Let us remind  that equations (1) have four integrals of motion
\be H_0 = l_{12}\,l_{34}+l_{23}\,l_{14}+l_{31}\,l_{24}=h_0,\ee
 \be H_1 = \sum _{j<k}^4 l_{jk}^2=h_1,\,\,
H_2 =\sum _{j<k}^4 (a_j+a_k)l_{jk}^2=h_2, \,\,
 H_3 = \sum_{j<k}^4 a_j a_k l_{jk}^2=h_3.\ee

Note that  $H_0$ and $H_1$ are the Casimir functions of
$so(4)$-Poisson structure, and the manifold ${\cal M}_h$ defined
by equations (6) -- (7)   is an affine part of two-dimensional
Abelian manifold (see Appendix by Mumford to the paper [AM
1982])\,\footnote{\,I am grateful to A. N. Tyurin for the
explanation
 of algebraic geometry related to this Appendix.}. Then formula (5) defines
 Hamiltonian vector field on ${\cal M}_h$.

The main result of this note is the following one: by elementary
means, it is shown that the dynamical variables $l_{jk}(t)$ are
expressed in terms of Abelian functions $f_{j4}(u_1,u_2)$,
$f_{kl}(u_1,u_2)$, $f_0(u_1,u_2)$, and $g(u_1,u_2)$ related to
genus two algebraic curve
 \be y^2=\prod _{j=0}^4 (x-d_j),\qquad d_0=0,\qquad d_4=d_1d_2d_3, \ee
with arguments depending linearly on time.

\medskip\noindent {\bf Theorem}. {\em Solution of equations} (1)
{\em has the form}
 \begin{eqnarray}
 m_j&=&l_{kl}=g(u_1,u_2)\,(\alpha _j\,f_{kl}(u_1,u_2)+\beta _j\,f_{j4}(u_1,u_2)),\\
 n_j &=&l_{j4} = g(u_1,u_2)\,(\gamma _j\,f_{kl}(u_1,u_2)+\delta _j\,f_{j4}(u_1,u_2)).
\end{eqnarray}
 {\em Here} $(j,k,l)$ {\em is a cyclic permutation of} (1,2,3),
 $\alpha _j$, $\beta _j$, $\gamma _j$, $\delta _j$,  {\em and} $d_j$
{\em are algebraic functions of integrals of motion and
quantities} $a_j$ {\em and} $b_k$. {\em Explicit expressions for
 them are given by} (24)--(26), (34), (35), (41), {\em and} (44).

\medskip\noindent
{\em Proof.} The key problem is the "uniformization" of the
manifold ${\cal M}_h$, i.e., finding of the "good" coordinates on
it. The proof consists of several steps.

\medskip\noindent
{\bf A.} Following K\"otter [Koe 1892] and using the linear change
of variables ${\em m_j}$ and ${\em n_j}$ to new variables
${\xi_j}$ and ${\eta_j}$, we transform equation (7) to the more
appropriate form:
  \be
 \sum _{j=1}^3\left ( \xi _j^2+\eta _j^2\right ) =0,\qquad
 \sum _{j=1}^3\xi _j\eta _j=0,\qquad \sum _{j=1}^3\left ( d_j\xi
_j^2+d_j^{-1}\eta _j^{2}\right ) =0. \ee
 For this, following Schottky [Sch 1891], let us introduce
 the three-dimensional vector ${\bf l}(s)$ depending on parameter $s$:
 \be
 {\bf l}(s) =(l_1(s), l_2(s),l_3(s)),\qquad
 l_j(s)=\sqrt{s_{j4}}\,m_{j} +\sqrt{s_{kl}}\,n_{j}, \ee
 where
 \be m_j=l_{kl},\qquad n_j=l_{j4},\qquad s_{jk}= (s-a_j)(s-a_k), \ee
 and $\{ j,k,l \}$ is a cyclic permutation of $\{1,2,3\}$. It is easy to
check that the function
\be
 f(s)={\bf l}(s)^2=\sum_{j=1}^3l_j(s)\,l_j(s) \ee
does not depend on time. So, it is the generating function of
integrals of motion
 \be
 f(s)=h_1s^2-h_2s+h_3+2\,h_0\sqrt{G(s)},\qquad G(s)=\prod _{j=1}^4 (s-a_j). \ee
From  formulae (12) and (14) it is easy to get the Lax
representation\,\footnote{\,However, this representation does not
need for the proof of Theorem. For the generalization of such
representation for the $n$-dimensional case see [Fe 2000].}
 \be \dot L(s) =[L(s), M(s)], \ee
 where $L(s)$ and $M(s)$ are antisymmetric matrices of the third order
corresponding to vectors
${\bf l}(s)$ and ${\bf m}(s)$,
\be
 {\bf m}(s)=(m_1(s),m_2(s),m_3(s)),\qquad m_j(s)=\sqrt{s_{kl}}\,m_j+\sqrt{s_{j4}}\,n_j,
\ee
 \be L(s)=\left( \begin{array}{ccc} 0&l_3 &-l_2 \\-l_3&0 &l_1\\
l_2 & -l_1& 0 \end{array} \right),\qquad M(s)=\left( \begin{array}
{ccc} 0& m_3 &-m_2\\ -m_3 & 0& m_1\\  m_2 &-m_1 & 0\end{array}
\right) .\ee

 The equation $f(s)=0$ is equivalent to the algebraic
equation of fourth degree $F(s) =\prod _{j=1}^4 (s-s_j)=0$,
 where
 \be F(s)= \left[ \left( h_1s^2-h_2s+h_3\right)
^2-4\,h_0^2\, G(s)\right] /\left( h_1^2-4h_0^2\right) .  \ee

This equation has four roots $s_1$, $s_2$, $s_3$, and $s_4$ that,
 in general, are complex ones. To them correspond four complex vectors
 \be {\bf l}^{(p)}={\bf l}(s_p)/\sqrt{F'(s_p)},\qquad p=1,2,3,4,\ee
(here $F'(s)$ is the derivative of $F(s)$) but only two of them,
for example ${\bf l}^{(1)}$ and ${\bf l}^{(2)}$, are linearly
independent, and
 \be \left( {\bf l}^{(p)}\right)^2=\sum_{k=1}^3\left( l_k^{(p)}\right)
 ^2=0,\,\,p=1,2,3,4;\,\, \sum _{p=1}^4\left( l_k^{(p)}\right)^2=0,\,\, k=1,2,3.\ee
 Let us introduce also the vectors $\xi$ and $\eta$ by the
 formulae 
\footnote{\, As it was noted by Yu. N.Fedorov, there is relation 
of these vectors to the problem
 of geodesics on two-dimensional ellipsoid with half-axes $\sqrt d_j$, j=1,2,3.
 Namely, $\xi$ may be considered as a tangent vector to geodesics 
and $i\eta$ as a normal vector
 to this geodesics.}
 \be \xi_j=l_j^{(1)}+i l_j^{(2)} ,\qquad \eta_j=l_j^{(1)}-il_j^{(2)}.\ee

Using (12) and (22) we may express $m_j$ and $n_j$ in terms of
$\xi_j$ and $\eta_j$
 \be m_j=\alpha _j \xi_j +\beta _j \eta_j,\qquad n_j=\gamma_j
 \xi_j+\delta _j\eta_j, \ee
 where
 \be
 \alpha_j =
 \frac{\sqrt{s_{kl}^{(2)}/F'(s_2)}-i\sqrt{s_{kl}^{(1)}/F'(s_1)}}{\Delta_j^{(3)}},\quad
 \beta_j = \frac{\sqrt{s_{kl}^{(2)}/F'(s_2)}+i\sqrt{s_{kl}^{(1)}/F'(s_1)}}{\Delta
 _j^{(3)}},\ee
 \[ \gamma _j =\frac{\sqrt{s_{j4}^{(2)}/F'(s_2)}-i\sqrt{s_{j4}^{(1)}/F'(s_1)}}{\Delta
 _j^{(3)}},\quad
 \delta _j = \frac{\sqrt{s_{j4}^{(2)}/F'(s_2)}+i\sqrt{s_{j4}^{(1)}/F'(s_1)}}{\Delta
 _j^{(3)}},\]
 \be \Delta _j^{(p)}=\frac{\sqrt{s_{j4}^{(q)}s_{kl}^{(r)}-s_{j4}^{(r)}s_{kl}^{(q)}}}
{\sqrt{F'(s_q)\,F'(s_r)}}.
\ee
 Here $(j, k, l)$ and $(p,q,r)$ are  cyclic permutations of (1, 2, 3).

Now it is easy to check that equations (7) take the form of three
K\"otter's quadrics (11), where
 \be \sqrt{d_j}=\frac{\Delta _j^{(1)}-i\Delta _j^{(2)}}{\Delta _j^{(3)}},\qquad
 \frac1{\sqrt{d_j}}=-\frac{\Delta _j^{(1)}+i\Delta _j^{(2)}}{\Delta _j^{(3)}}.\ee

\medskip\noindent
{\bf B.} Following  [Koe 1892], let us show that the manifold
defined by equations (11) may be "uniformized" by means of the
 Weierstrass Wurzelfunctionen related to the hyper-elliptic curve
 (8) that are defined as
\begin{eqnarray}
P_j(z_1,z_2) &=& \sqrt{(z_1-d_j)(z_2-d_j)},\qquad j,k=0,1,2,3,4, \\
 P_{jk}(z_1,z_2) &=& \frac{P_j P_k}{(z_1-z_2)}\left( \frac{\sqrt{R(z_1)}}
{(z_1-d_j) (z_1-d_k)}-
 \frac{\sqrt{R(z_2)}}{(z_2-d_j)(z_2-d_k)}\right) .
 \end{eqnarray}

These sixteen functions $P_j(z_1,z_2)$ and $P_{jk}(z_1,z_2)$
satisfy a lot of identities. All of them  may be
 obtained from  definitions (27) and (28)
 (for details, see [We 1878], [Koe 1892], and [Ca 1893]).
 \footnote{See also modern survey [BEL 1997].}
 Here we give only few of them which are useful for us:
\be
 \sum _{j=1}^3 c_j\left(\frac{P_{kl}^2}{(s-d_k)(s-d_l)}
 +\frac{P_{j4}^2}{(s-d_j)(s-d_4)}\right)=
 \frac{s}{\prod _{j=1}^4(s-d_j)},\ee
 \be \sum _{j=1}^3 {\tilde c}_j\,P_{j4}^2=d_4,\qquad
 \sum _{j=1}^3d_j{\tilde c}_j\,P_{kl}^2=P_0^2,\ee
 \be \sum _{j=1}^3c_j\,P_{j4}P_{kl}=0,
\qquad \sum _{j=1}^3{\tilde c}_j\,P_{j4}P_{kl}=-P_0,\ee
 \be \sum _{j=1}^3 c_j\left( d_j^{-1}P_{j4}^2+d_jP_{kl}^2\right) =0, \ee
 where
 \be {\tilde c}_j=\frac1{(d_j-d_k)(d_j-d_l)},\qquad c_j
 =\frac{d_j-d_4}{(d_j-d_k)(d_j-d_l)}.\ee

It is known (see [We 1878]) that $P_j(z_1,z_2)$ and
$P_{jk}(z_1,z_2)$ up to the factors are the ratio of the theta
functions with half-integer theta characteristics\, \footnote{\,We
give here just one series of such expressions. Relative other
series, see [Koe 1892].}
 \be
 P_j(z_1,z_2)=f_j(u_1,u_2)=\frac{\theta _j(u_1,u_2)}{\theta (u_1,u_2)},\quad
 P_{kl}(z_1,z_2)=f_{kl}(u_1,u_2)=\frac{\theta _{kl}(u_1,u_2)}
 {\theta (u_1,u_2)}\,, \ee
\begin{eqnarray}
\theta _{23}(u_1,u_2) &=& \theta \left[ \begin{array}{cc} 0&0\\
1&1
\end{array}\right] (u_1,u_2),\qquad \theta _{31}(u_1,u_2)=\theta \left[
\begin{array}{cc}1&0 \\ 1&1\end{array} \right] (u_1,u_2),\nonumber \\
\theta _{12}(u_1,u_2) &=& \theta \left[ \begin{array}{cc}1&0\\ 0&0
\end{array} \right] (u_1,u_2),\qquad \theta _{14}(u_1,u_2) =
\theta \left[ \begin{array}{cc} 1&1\\ 1&1 \end{array}\right]
(u_1,u_2),\nonumber \\
\theta _{24}(u_1,u_2) &=& \theta \left[ \begin{array}{cc}0&1 \\
1&1
\end{array} \right] (u_1,u_2),\qquad \theta _{34}(u_1,u_2) = \theta \left[
\begin{array}{cc}0&1\\ 0&0\end{array}\right] (u_1,u_2),\nonumber \\
\theta _{0}(u_1,u_2) &=& \theta \left[ \begin{array}{cc} 1&1\\ 0&0
\end{array}\right] (u_1,u_2),\qquad \theta (u_1,u_2)=\theta \left[
\begin{array}{cc}0&0 \\ 0&0\end{array} \right] (u_1,u_2).
\nonumber \\
 &&\end{eqnarray}
Here
 \be \theta (u_1,u_2)=\sum _{n_1,n_2} \exp\{ i\pi (n_1(2u_1+n_1\tau _{11}+
n_2\tau _{12})+n_2(2u_2+n_1\tau _{21}+n_2\tau _{22}))\}, \ee
 and $\tau _{jk}$ are elements of period matrix.

 The comparison (11) with (29)--(33) shows that
 \be \xi_j=\sqrt{c_j}g \,P_{kl},\qquad \eta_j=\sqrt{c_j} g \,P_{j4},\ee
  where $g$ is an unknown function.

\medskip\noindent
 {\bf C.} The rest part of the proof is the uniformization of equation (6).

Let us substitute expressions (23)  for $m_j$ and $n_j$ into
equation (6). Then by using of (24) and (25) we transform it to
the form
 \be H_0=\sum _{j=1}^3 \left( A_j\left( \xi _j^2-\eta _j^2\right )+
 B_j\xi _j\eta _j\right) =h_0,\ee
where
 \be
 A_j=\alpha +\beta d_j+\gamma {d_j}^{-1},\qquad B_j=\delta \left( d_j+d_j^{-1}\right ).
\ee
Here $\alpha , \beta , \gamma $, and $\delta $ are algebraic
functions of $h_0, h_1, h_2, h_3, a_j$, and $b_j$.

This sum may be calculated by using of (24), (29)--(33). The
result is
 \be H_0=\frac{(1-\varepsilon P_0)^2}{4\varepsilon d_4}\,g^2 =h_0,\ee
where
 \be \varepsilon=\frac{\sqrt{d_4}\left( \sqrt{(s_3-s_1)(s_2-s_4)}
-\sqrt{(s_2-s_3)(s_1-s_4)}\right)}
 {\sqrt{(s_1-s_2)(s_3-s_4)}}. \ee

From this we obtain
 \be g=(1-\varepsilon f_0)^{-1},\qquad \xi _j=g\sqrt{c_{j}}\,f_{kl},\qquad
 \eta _j=g\sqrt{c_{j}}\,f_{j4}.\ee
The fact of linear dependence of arguments $u_1$ and $u_2$ on time
$t$ follows as from the algebraic geometrical approach [AM 1982]
as from the old K\"otter approach [Koe 1892].

This completes the proof of Theorem.

\medskip\noindent
{\bf Acknowledgements}. The main part of this paper was completed
during my visit to Max-Planck Institute for Mathematics, Bonn. I
am grateful to the stuff of this Institute and also to the
Department of Theoretical Physics of Zaragoza University for
hospitality.


\begin{thebibliography}{lll}

\bibitem[AM 1982]{AM1} Adler M. and van Moerbeke P., {\em The algebraic
integrability of geodesic flow on} $SO(4)$, Inv. Math. {\bf 67},
297--326; with appendix by D. Mumford, pp. 326 -- 331

\bibitem[AM 1988]{AM2} Adler M. and van Moerbeke P., {\em The Kowalewski and
H\'enon-Heiles motions as Manakov geodesic flows on $SO(4)$ -- a
two-dimensional family of Lax pairs}, CMP {\bf 113}, 659--700

\bibitem[BEL 1997]{BEL} Bukhstaber V.M., Enolskii V.Z. and Leikin D.V.
{\em Hyperelliptic Kleinian functions and applications}, Solitons,
geometry, and topology: on crossroad, 1-33, Amer. Math. Soc.
Transl. Ser 2, 179, Amer. Math. Soc., Providence, RI

\bibitem[Bo 1986]{Bo} Bobenko A.I., {\em Euler equations on the Lie algebras
$e(3)$ and $so(4)$. Isomorphism of integrable cases}, Funct. Anal.
Appl. {\bf 20}, 53--56

\bibitem[Ca 1893]{CA} Caspary F., {\em Sur une nouvelle mani\`ere d'\'etablir
les relations alg\'ebriques qui ont lieu entre les fonctions
hyperelliptiques de premi\`ere esp\`ece}, Ann. Sci. Ec. Norm.
Sup., 3 Ser., {\bf 10}, 253--294

\bibitem[Ca 1846]{Ca} Cayley A., {\em Sur quelques propri\'et\'es des
d\'et\'erminants gauches}, J. Reine Angew. Math. {\bf 32}, 119-123

\bibitem[Cl 1871]{Cl} Clebsch A., {\em \"Uber die Bewegung eines K\"orpers
in einer Fl\"ussigkeit}, Math. Ann. {\bf 3}, 238-261

\bibitem[Fe 2000]{FE2} Fedorov Yu.N., {\em Integrable systems, Poisson
pencils, and hyperelliptic Lax pairs}, Reg. Chaot. Dyn. {\bf 5},
No.2, 171--180

\bibitem[Fr 1874]{Fr} Frahm F., {\em \"Uber gewisse Differentialgleichungen},
Math. Ann. {\bf 8}, 35--44

\bibitem[Ha 1983]{Ha} Haine L., {\em Geodesic flow on $SO(4)$ and abelian
surfaces}, Math. Ann. {\bf 263}, No.4, 435--472

\bibitem[Koe 1892]{Ko} K\"otter F., {\em \"Uber die Bewegung eines festen
K\"orpers in einer Fl\"ussigkeit}, I, II, J. Reine Angew. Math.
{\bf 109}, 51--81, 89--111

\bibitem[Ma 1977]{Ma} Manakov S.V., {\em Note on the integration of
Euler's equations of the dynamics of an $n$-dimensional rigid
body}, Funct. Anal. Appl. {\bf 10}, 328--329

\bibitem[Mo 1980]{MOS} Moser Ju., {\em Geometry of quadrics and spectral
theory}, in: {\em Chern Symposium 1979}, 147-188

\bibitem[Sch 1891]{Sch} Schottky F., {\em \"Uber das analytische Problem der
Rotation eines starren K\"orpers im Raume von vier Dimensionen},
\b Sitzungber. K\"onig. Preuss. Akad. Wiss. zu Berlin, 227--232

\bibitem[We 1878]{We} Weber H., {\em Anwendung der Thetafunctionen zweiter
Ver\"anderlich-}\b {\em er auf die Theorie der Bewegung eines
festen K\"orpers in einer Fl\"ussig-}\b {\em keit}, Math. Ann.
{\bf 14}, 173--206

\end{thebibliography}
\end{document}